# Friction Force for Self-Excited Systems

Mario D'Acunto

*Department of Earth Science, University of Pisa, via S. Maria 53, I-56100 Pisa, Italy*

**Abstract.**
Control and reduction of friction is one of the last frontier in the field of tribology. This paper presents a model for low friction made up of a self-excited oscillator sliding along a periodic potential. Many examples of such self-excited oscillators in electrical, mechanical and biological systems can be found in literature. As one of its most prominent features, stable limit cycles in phase space are established, emerging from a balance between energy gain and dissipation. The possible role of self-excitation on friction force is based on the idea that a limit cycle represents a no dissipation dynamics. If a self-excited oscillator is pulled by a stage and it starts to slide, then, because the variation of position, the sliding process destroys the limit cycle setting up a stick-slip dynamics. Nevertheless, a reduction of energy could be observed if the self-excitation restoring the limit cycle is not completely destroyed by the sliding movement. It will be demonstrated that a self-excited sliding body tangential to a surface experiments changes of stick-slip dynamics and a low reduction of energy dissipation during the motion.

1. **Introduction**

When two unlubricated solid bodies contact each other and one body subsequently slides against the other, a frictional force occurs. Several models have been proposed to explain the origin of this frictional force in macroscopic or in microscopic systems [1-9].
In the studies of friction one possible starting point is the equation of motion

$$m\ddot{x} + \Phi(x,\mu)\dot{x} = -\frac{\partial U}{\partial x} + f + \eta \qquad (1)$$

where $x$ is the coordinate of the oscillator, $m$ is the mass, $f$ is a driving force, $\eta$ is the noise, $U(x)$ is an external periodic potential and $\Phi(x,\mu)$ is the damping function.
On the microscopic scale, it is possible to consider conditions in which the energy dissipation is reduced until a nearly frictionless motion condition is achieved [10-14]. Starting from the pioneering model for wearless friction of Tomlinson [15], recent studies modeled the occurrence of friction as a consequence of the existence of unstable equilibrium positions during the translations of one body over another. Moving through these equilibrium positions leads to sudden displacement, with the consequence that sliding body vibrates and this vibration excites the bulk in an irreversible manner, causing loss in energy [16-17].
Dissipation processes are described phenomenologically, in classical equations of motions, by a damping term. This term is usually assumed to be proportional to velocity, i.e., the damping function is constant $\Phi(x,\mu)=\mu$, the so-called Coulomb damping [18-19]. In this paper, it will be considered a mechanism for which the energy lost during the motion can be reintroduced in the energetic balance of the sliding body. Such mechanism can be analytically condensed in eq. (1) by adopting

$$\Phi(x,\mu) = \mu(x^2-1) \qquad (2)$$

for the damping term, i.e., a damping coefficient dependent both on velocity and position. A damping function as expressed in equation (2) was introduced firstly by B. Van der Pol in order to describe a self-excited oscillator [20]. A self-excited oscillator is a well known system which has some external source of energy upon which it can draw [18-20]. The *Van der Pol* oscillator is, probably, the most simple example of such system. As one of its most prominent features, stable limit cycles in phase space are established, emerging from a balance between energy gain and dissipation. The possible influence of self-excitation dynamics on friction force is based on the idea that a limit cycle represents a no dissipation dynamics. Let us consider a body initially experimenting a limit cycle, its position and velocity are well established in the phase space. If the body is pulled by a stage moving on a straight line, than the body could travel with a reduction of energy if the self-excited damping function restoring the limit cycle is not completely destroyed by the sliding movement. It will be demonstrated that a self-excited sliding body tangential to a surface experiments changes of stick-slip dynamics and reduction of energy dissipation during the motion.

## 2. The model

A single self-excited sliding body tangential to a surface can be built in different ways, [14-16]. Figure 1 gives a schematic sketch of a self-excitated oscillator as considered in this paper.

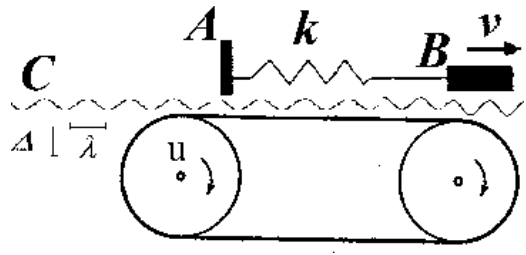

Figure 1. Schematic sketch of a self-excited oscillator (*A*) sliding on a periodic potential. The upper body of interest *A* is pulled by a linear spring with a constant force connected to a stage (*B*) that moves with constant velocity *v*. The lower body (C) is a rough surface moved by a belt rotating with lower angular velocity *u* respect to the stage. The rough surface is simulated by a periodic potential, $\Delta$ is the strength of periodic potential, λ its period. All parameters and variables are given in dimensionless units.

The body of interest (upper sliding body) *A* is elastically coupled to a large mass *B* traveling on a straight line with constant velocity *v* and moves along the direction of the velocity. The surface of the belt *C* (lower body) is rough, and the roughness is described by a periodic potential. While stage moves steadily, the body of interest drops periodically into a potential minimum. As a consequence, it is dragged out again. The motion of the stage can be arbitrarily slow, and yet the oscillator performs a fast motion while it is falling. After a dropping motion the body *A* is again at a minimum of potential.

The equation of motion for a unit mass self-excited damping function oscillator, SEDF, can be described as follows:

$$\ddot{x} + \mu(\beta x^2 - 1)\dot{x} + \Delta sin(2\pi x/\lambda) + \Omega^2(x - vt) = 0 \qquad (3)$$

where $\mu$ is the damping coefficient, $\Delta$ is the strength of periodic potential, $\lambda$ is the period of the potential, $\Omega$ is the natural frequency of the oscillator, and *v* is the external driving velocity. The coefficient $0 \leq \beta \leq 1$ is introduced in the equation of motion (3) in order to quantify the self-excitability of the system. If $v=0$, $\Delta=0$, and $\beta=1$, equation (3) is reduced to a standard *Van der Pol* oscillator. The external potential is

$$U(x,t) = \frac{1}{2}\Omega^2(x-vt)^2 - \frac{\lambda}{2\pi}\Delta\cos(2\pi x/\lambda) \qquad (4).$$

It presents an absolute minimum point in $x=0$, and a series of periodic minima, figure 2.

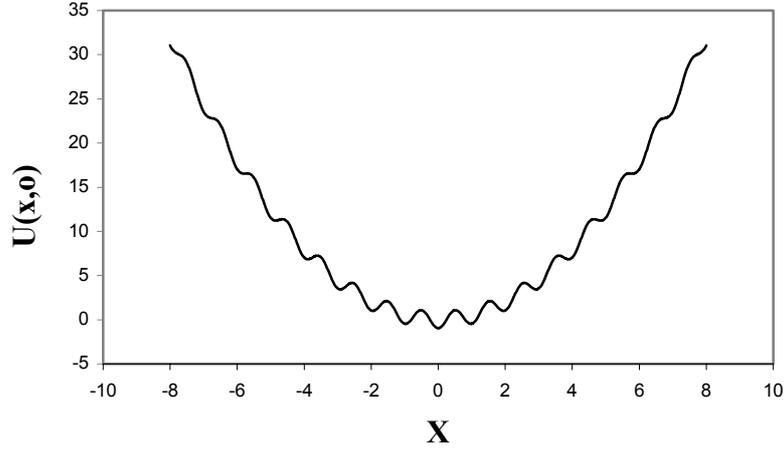

Figure 2. Potential $U(x,0)$ versus position, $\lambda=1$.

When $v=0$, the oscillator reaches to limit cycle (steady state) which is weakly or strongly deformed as a function of the strength of potential $\Delta$ and coefficient $\beta$. During the motion the dynamic friction force may either introduce energy into the system, if it has the same sign as the velocity of the body A, or dissipate energy, if the friction force and the velocity of the body A have different signs. It is possible to show by analytically that during a limit cycle the energy dissipation is zero. Adopting the Krylov-Bogoliubov-Mitropolsky (KBM) averaging method [20], the displacement of the position of the second approximation can be written as follows

$$x = a\cos\psi + \mu f(a,\psi) + \mu^2 g(a,\psi) \qquad (5)$$

where

$$\psi = \left[1 - \mu^2(1-\beta a^2/4)(1/3 - \beta a^2/4)\right]t + \psi_0 \qquad (6)$$

$$\dot{a} = \mu\left[\frac{a}{2}(1-\beta a^2/4) + \frac{\Delta}{\mu}J_1(ka)\right] + \mu^2\left[\frac{k\Delta^2}{2\mu}J_0(ka)J_1(ka)\right] \qquad (7)$$

$$f(a,\psi) = -\frac{\beta a^3}{32}\sin 3\psi + \frac{\Delta}{\mu}\sum_{n\neq 1}(-1)^n J_{2n+1}(ka)\cos((2n+1)\psi) \qquad (8)$$

$$g(a,\psi) = -\frac{k\Delta^2}{\mu}J_0(ka)J_1(ka) - 2a(1-\beta a^2/4)(1/3 - \beta a^2/4)\cos\psi \qquad (9)$$

where $k=2\pi/\lambda$, and $J_0$, $J_1$, $J_{2n+1}$ are Bessel functions of first kind. For $\Delta\neq 0$ and $v=0$, SEDF described by eq. (3) shows limit cycles. Examples of such limit cycles for several $\beta$ parameter are plotted in Figure 3.

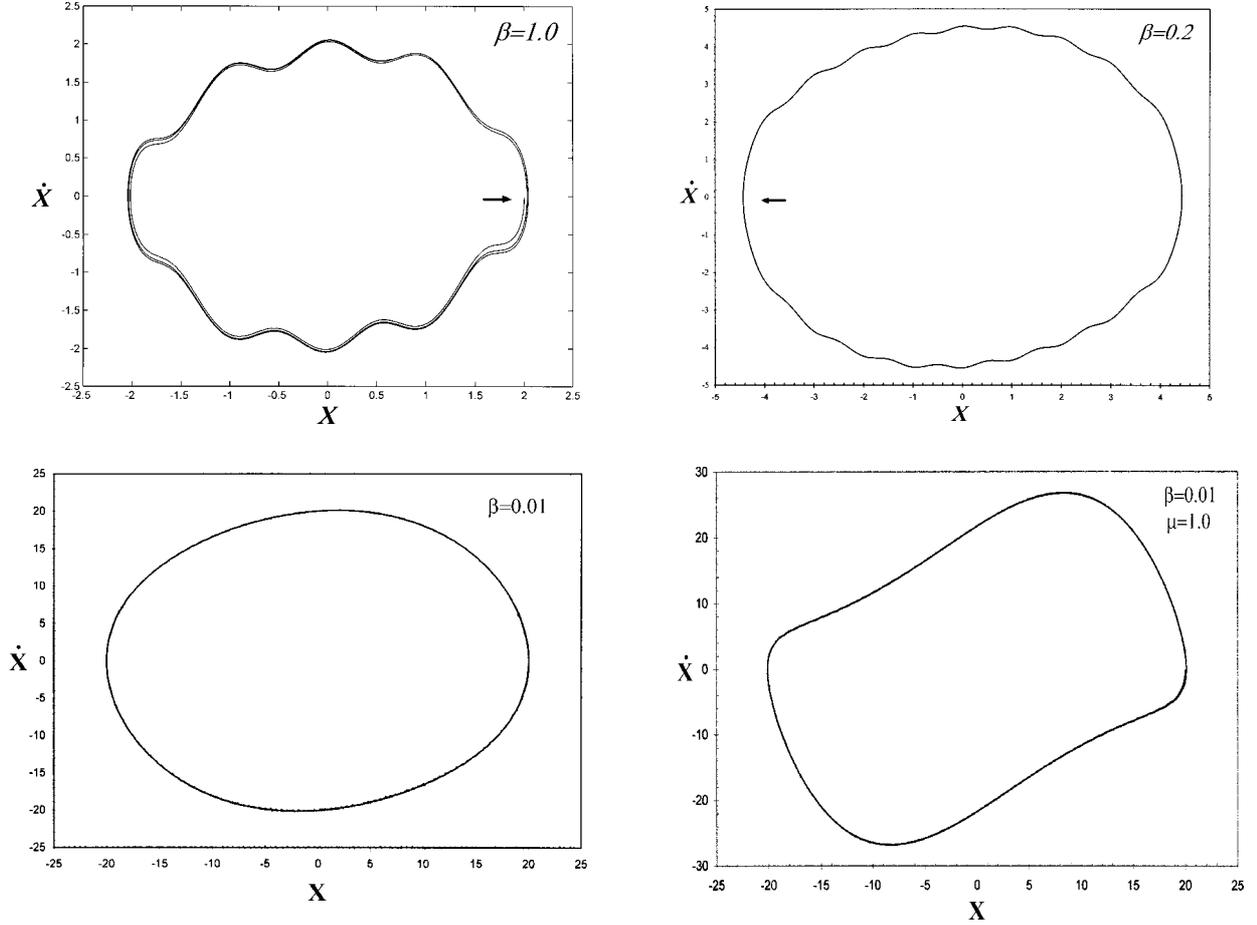

Figure 3. Phase space plots for undriven SEDF, $\Delta=1.5$. The arrow displays the initial coordinate.

If the non-autonomous case is bring into focus, i.e., $v\neq 0$, the system can drop between potential well. Friction force is produced during the transition between potential wells, and it shows the characteristic stick-slip behavior. During the motion, the energy gain and loss at the time $\tau$ can be expressed by the relation:

$$\Gamma = \mu \int_0^\tau (\beta x^2 - 1)\dot{x}^2 dt \qquad (10)$$

The expression (10) can assume negative values, it means that the system slides and absorbs energy from the external source. For this reason, it is useful to split the dissipation function in two terms:

$$\Gamma = \beta\Gamma_+ - \Gamma_- \qquad (11)$$

where the signs + and - are referred to positive or negative contribution, respectively. The range of interest, i.e., the spatial region where the self-excitation effects the dynamics, is given by the condition $\beta\Gamma_+\approx\Gamma_-$. For large values of the position, i.e., $|x|\gg 1/\sqrt{\beta}$, energy dissipation is entirely positive $\Gamma \cong \beta\Gamma_+$, and the self-excitation contribution disappears. The evaluation of dissipation energy during a limit cycle requires the integration of eq. (10). The integration of eq. (10) involves eq. (5) and its time derivative, that is obtained with the aid of equations (6-9). The integral of dissipation evaluated during a period results to be nearly zero. The period can be calculated numerically.

Recently, frictional models based on Coulomb damping has been intensively studied. The equation which describes the model can be obtained by eq. (3) when $\beta=0$, and $\mu \rightarrow -\mu$

$$\ddot{x} + \mu \dot{x} + \Delta sin(2\pi x/\lambda) + \Omega^2 (x-vt) = 0 \qquad (12)$$

The dynamics of SEDF will be compared with that obtained by eq. (12). For sake of simplicity, we will refer the sliding body described by eq. (12) as Coulomb damping constant, CDC, sliding body. The dynamics of CDC can be found in literature [23-24]. CDC delineates both dry friction or frictional forces in a thin layer of electrolyte solution confined between two plates. The basic mechanism of stick-slip dynamics for a CDC can be explained as follows. If the velocity of the stage is small, $v \ll 1$, the motion involves two steps: slow motion in a local minima of the total potential $U(x)$, and a fast slip (sliding) that begins when an instability occurs, i.e., $d^2U/dx^2$ changes sign. At the point of instability the spring force, $\Omega^2(vt-x)$, reaches a maximum value corresponding to the static friction force. The static friction force is generally equals to the strength of potential, $\Delta$. During a slip the spring force decreases until it reaches a value where the sliding ceases and the top plate is trapped again at a potential minima. Then a periodic stick-slip motion of the upper body A, Figure 1, is observed. The dynamics of the stick-slip motion can be analyzed taking into account that the stage is effectively at rest during the fast slip of the stage, $vt=L_0$=const. The slip motion corresponds to the jump between nearest-neighbor minima of the potential $U(x,L_0)$, located at $x_0$ and $x_1$ respectively. The energy dissipated during the slip is given by $\Delta W = U(x_0,L_0) - U(x_1,L_0)$ where $x_0$ and $L_0$ are the position of the body $A$ and the length of the spring at the saddle point given by $dU/dx=0$ and $d^2U/dx^2=0$.

When the stage ($B$) in Figure 1 is at the rest, a limit cycle represents a no dissipation dynamics of the SEDF. The self-excitated oscillator drops between potential well balancing energy gain and dissipation, the main result is a stationary oscillation of position and velocity as visible in Figure 3. The main question is: what happen if the SEDF, initially located on a limit cycle starts to slide? Two mean situations are possible. The first one, if the limit cycle is destroyed then the no dissipation dynamics is lost and dissipative motion takes place. Second one, the conditions restoring a limit cycle permits to balance dissipation during the sliding motion. In this case, the SEDF should slide losing smaller amount of energy. The paper will show that in the spatial range where the self-excitation is active, i.e., $|x| \leq 1/\sqrt{\beta}$, the SEDF dissipation energy during sliding motion is less respect CDC dissipation energy. Outside of this range, the frictional term in eq. (3) changes as $\approx \mu x^2 \dot{x}$, overcoming self-excitation and enhancing kinetic friction.

3. **Numerical Results and Discussion**

The main goal of this paragraph is to obtain some meaningful results via a phenomenological approach. Eq. (3) has been integrated numerically using a fourth-order Runge-Kutta algorithm with adaptive step size [25], and with parameter values fixed at $\mu$=0.1 and $\Omega$=1.0. The purpose of the investigation is to give prominence to the persistence of self-excitation when a sliding process takes place.

Figure 4 displays characteristic motion in the phase space and series with typical stick-slip dynamics. It is remarkable the competition between pinning action of potential wells versus driven force.

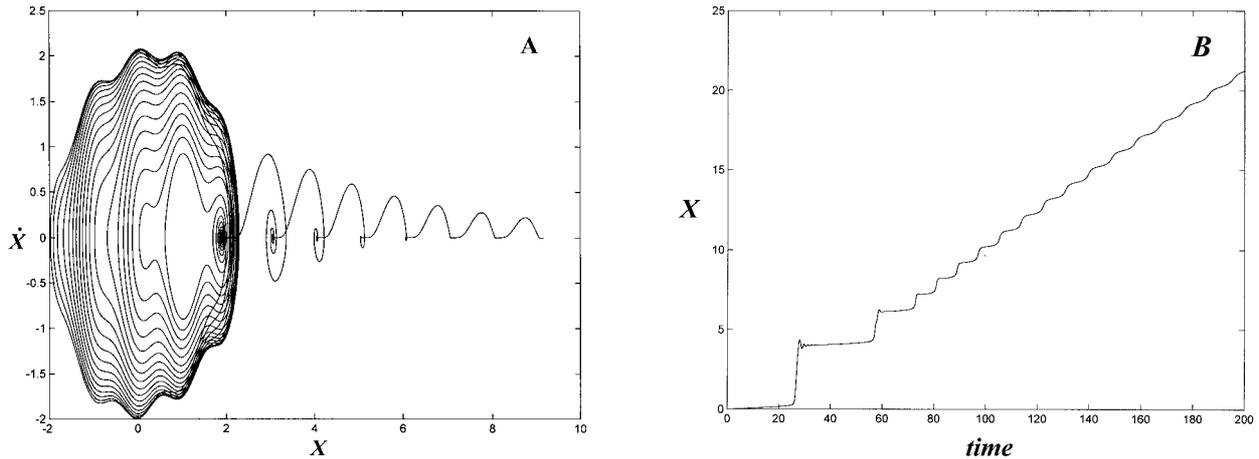

Figure 4. Phase space (A) and series plots (B) showing the characteristic sliding process for SEDF, $\Delta=3.0; \beta=1.0$; and $v=0.1$. Figure 4,A shows the deformation of a limit cycle due to the spring force. The initial condition for the series is the origin.

Because the different influence of self-exciting damping function $\mu(\beta x^2-1)$, the dynamics is strictly divided in two separate regimes. The first one is dominated by the fact that the self-excitation damping function affects the sliding motion of the SEDF. The spatial range of this regime is given by the relation $|x|\leq 1/\sqrt{\beta}$. The pulling force deforms the stationary trajectory of the SEDF, located initially on a limit cycle.

In the second regime, the term $\beta x^2$ is more greater than the unity and the self-excitation is destroyed enhancing the dissipation during the sliding process. The motion is characterised by rapid drop of potential well and disappearance of stick-slip. The stick-slip mechanism seems to be very close to the pinning-depinning transition [5,8]. The parabolic form of the damping function in SEDF makes more drastic the separation between stick (pinning) and slip (sliding) phase. The damping function forces the body A to live in a potential well, so much that the spring term permits to drop in a subsequent potential well. Independently by the position assumed in the new location well, the SEDF experiments a reduced escape rate lowering the frictional force involved during the jump.
Figure 5 shows a comparison between CDC and SEDF dynamics.

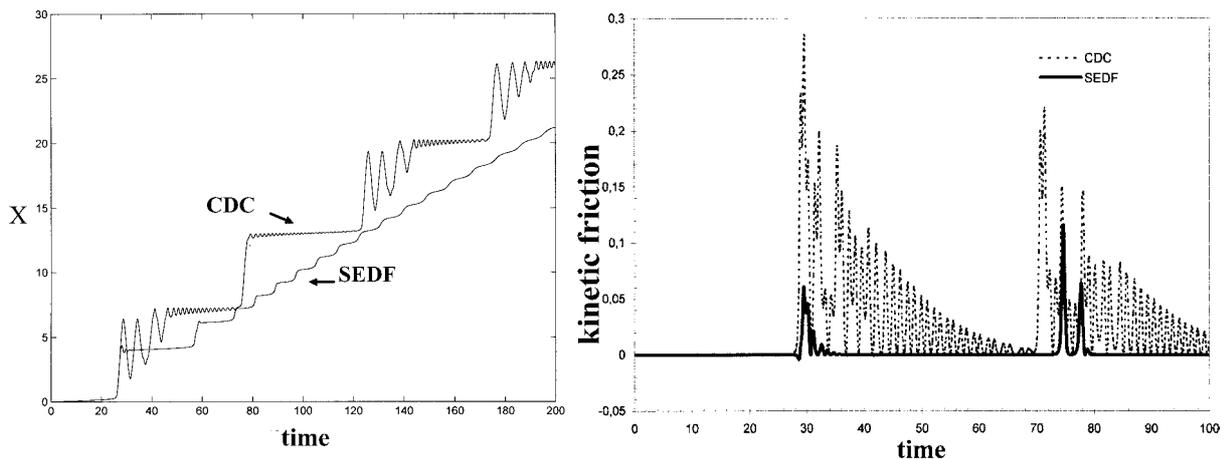

Figure 5. Time series comparison for SEDF (lower line), and CDC (upper line); $\Delta=3.0$, $\beta=1.0$, $v=0.01$.

Initially, the slip phase for the SEDF is represented by a single jump, while the CDC presents an oscillation before the stabilization. The peak of kinetic friction of SEDF is smaller of about 5 times respect to the correspondent CDC peak, Figure 5. Afterwards, reduction of stick-lip and increasing kinetic friction dominate the dynamics. This dynamics can be explained by the following reasoning.

The condition for a stick solution is given by the relation $\mu\beta x^2 \dot{x} + vt \approx \partial U / \partial x + \mu\dot{x}$, where $U$ is the potential shown in Figure 2. The self-excitability coefficient $\beta$ introduces negative damping that balances the constant frictional term. As a consequence, if the strength of the periodic potential $\Delta$ is decreased, then the time needed to start for stick-slip is equally decreased. For $|x-vt|>\Delta/\Omega^2$ it is impossible to have stick solutions. Initially, lying in a potential well, the oscillator vibrates until a saddle-node bifurcation is reached where a stable state annihilates in an unstable one. Then, the position changes as $x \approx vt + \Delta \sin 2\pi x$, Figure 6. As a function of the stage $B$ velocity, the body $A$ experiments decreasing oscillations between subsequent potential wells and the oscillations are driven by the sinusoidal term deriving by the potential $U$. Outside of the self-excitation range, because the modified damping term proportional to $\mu\beta x^2$, the distance between the body $A$ and the stage $B$, quantified by $|x-vt|$, increases. This increment is a consequence of the fact that the spring force, $\Omega^2(vt-x)$, does not compensate the growth of viscous component of friction [23]. A first consequence of this increment is that the body $A$ cannot fall in a potential well and the stick-slip dynamics disappears. Figure 6 shows the asymptotic behaviour of kinetic friction for a SEDF. Due to the viscous component of friction which increases as $\approx \mu x^2 \dot{x}$, the friction diverges outside the self-excitation range.

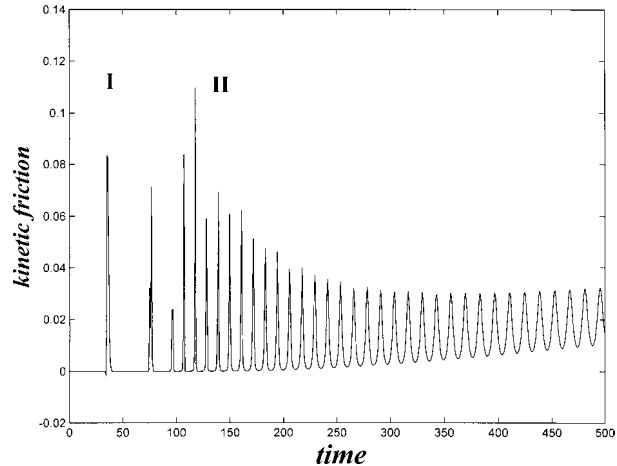

Figure 6. Kinetic friction for SEDF vs time. I indicates the self-excited region and II indicates the frictional increment due to viscous component. The plot shows the divergent behaviour of the friction when the self-excitation is lost.

A new class of results should be expected varying the several parameters involved in the equation of motion (3). In order to simplify the dynamics when affected by self-excitation, only significant selected values have been taken in consideration. A relevant property of the self-excited oscillator is the existence of limit cycle for small values of $\beta$. Figure 3 shows that the spatial range involved in a limit cycle for small value of $\beta$ is largely increased. This meaningful property could have a wide series of experimental applications in the field of tribology characterization of materials [2], as well as in applied mechanical needed.

4. **Conclusions**

In the last fifteen years the interest in friction and dissipative processes between solids has been stimulated due to the possibility of reproducible friction experiments on the mesoscopic and nanoscopic scale. There is some hope that such less complex systems an understanding is possible by theoretical models. Furthermore, there is also some hope that it should to be possible to develop simple models showing the basic phenomenological features of friction.

This work discusses a simple model of classical mechanics with the purpose of describing reduction of friction during a solid-solid sliding process. The model consists a self-excited oscillator sliding along a periodic potential and subject to a linear force. The results shown in Figure 3-6 are the mean contribution of the present paper in order to understanding the possible role of self-excitation mechanism on the study of frictional forces.

In order to give emphasis to the results on stick-slip dynamics and friction force, the dynamics of the self-excited oscillator (equation 3) has been compared with the correspondent features of Coulomb damping dependent oscillator (equation 12).